\documentclass[aps,twocolumn,pra,showpacs,floatfix]{revtex4}
\usepackage{epsfig}
\usepackage{graphicx}
\usepackage{dcolumn}
\usepackage{amsmath}

\begin{document}

\title{Axio-electric effect }

\author{A. Derevianko}
\affiliation {
Department of Physics, University of Nevada, Reno, Nevada 89557}
\author{V. A. Dzuba}
\affiliation{School of Physics, University of New South Wales, Sydney 2052, 
Australia}
\affiliation {
Department of Physics, University of Nevada, Reno, Nevada 89557}
\author{V. V. Flambaum}
\affiliation{School of Physics, University of New South Wales, Sydney 2052, 
Australia}
\author{M. Pospelov}
\affiliation{Department of Physics and Astronomy, University of Victoria, 
Victoria, British Colombia, V8P IAI, Canada}
\affiliation{Perimeter Institute for Theoretical Physics, Waterloo,
Ontario, N2J 2W9, Canada}

\date{\today}

\begin{abstract}
Using the relativistic Hartree-Fock approximation, we calculate 
the rates of atomic ionization by absorption of axions of the
energies up to 100 keV and for an arbitrary value of the axion mass.
We present numerical results for atoms used in the 
low radioactive background searches of dark matter ({\em e.g.} Ar, Ge and Xe),
as well as the analytical formula which fits numerical calculations 
for the absorption cross sections and can be applied for other atoms,
molecules and condensed matter systems. 
Using the cross-sections for the axio-electric effect, we derive the
counting rates  
induced by solar axions and set limits on the axion coupling constants.

\end{abstract}


\maketitle

\section{Introduction}

The idea of dynamical adjustment of the electroweak vacuum 
that cancels the $\theta$-angle of QCD \cite{PQ} is perhaps the most 
natural solution to the strong $CP$ problem. This mechanism 
inevitably leads to the conclusion about the existence of light pseudoscalar 
particle in the spectrum, called axion \cite{WW}. Breaking of axial
$U(1)$ symmetry by QCD anomaly  
gives a nonperturbative mass to axions with 
\begin{equation}
\label{mass}
m_a^2 \propto \frac{m_* |\langle \bar qq \rangle|}{f_a^2}.
\end{equation}
Here $m_* = (m_u^{-1} + m_d^{-1})^{-1}$ is the combination of quark masses, $\langle \bar qq \rangle$ 
is the quark condensate and $f_a$ is the axion coupling scale. 
While original models linked $f_a$ to the weak scale, it was soon realized that 
it can be in fact arbitrarily large \cite{invisibleaxions}, limited only by 
cosmological and astrophysical considerations (see, {\em e.g.} \cite{Peccei}). 

While the mass of the QCD axion is rigidly linked to its coupling with the 
topological term $G_{\mu\nu}^a \tilde G^a_{\mu\nu}$ via Eq. (\ref{mass}), any axion model 
allows for additional derivative type couplings to axial currents $ J^A_\mu \partial_\mu a/f_a $
of quarks and photons, that obey the shift symmetry of the axion interactions. 
Over the years a lot of experimental activity has been devoted to detecting 
axions using interactions of this form. Some methods employ 
finite cosmological number density of relic axions, while others use the idea of detecting axions that 
are produced in the solar interior. For a comprehensive review
of axion-related phenomenology, see, {\em e.g.} Refs. \cite{Turner,Sikivie,Raffelt}. 

A dedicated search for solar axions, such as CAST \cite{CAST}, 
uses conversion of keV-energy axions into x-ray photons in the magnetic field. 
Although stringent constraints on the axion coupling constant have been 
imposed by such searches, only recently did they become competitive with 
the broad range of astrophysical constraints.

An alternative way of detecting solar axions was proposed in
Refs. \cite{Avignone}. 
The coupling of axions to electrons can lead to the atomic ionization
and therefore be searched with  
high radio-purity materials in the underground experiments. 
Recent decade has seen a proliferation of such experiments, that
source their main scientific  
motivation in searching for the nuclear recoil from scattering of
weakly-interacting massive particles (WIMPs), 
a putative component of galactic dark matter. Many of these
experiments are also able to detect ionization created 
by solar axions down to a relevant energy scale of a few keV. Some
constraints on solar axions  
were already imposed by the CDMS experiment \cite{CDMS}.  This
analysis was also  
extended to the absorption of the super-weakly interacting 
massive particles (super-WIMPs) that may also plausibly be a dark matter 
candidate \cite{DAMA2}. In case of the pseudoscalar particles, the 
latter possibility departs, of course, from the
mass-coupling relation suggested by (\ref{mass}). To make the
distinction clear, we shall designate the solar axions as "massless"
or relativistic, and refer to the massive keV-scale axions as
super-WIMP possibility.  
The constraints on super-WIMP axions were improved recently in
Ref. \cite{CoGent}.  

Up until this year, the theory of axio-electric effect was using very
simplistic formulae  
relating the cross sections of axion absorption to the photo-electric
one \cite{Dimopoulos,Pospelov}.  
Earlier this year, the three of us have updated these calculation for
the case of the  
massive axions using the relativistic Hartree-Fock calculations
\cite{DzuFlaPos}.  
In this paper, we calculate the 
axio-electric effect caused by axions of arbitrary mass, including the
relativistic case. Convoluted with the  
flux of the axions emitted by the solar interior, these results would enable 
searching/setting limits on the models of light axions that 
have direct couplings to electrons. Such calculations are especially
timely in light of  several  
 dark matter experiments have reporting the excess of events over the 
expected background in the keV region \cite{CoGent} (see also
Ref. \cite{DAMA1}, where the  
annual modulation of the keV-scale energy deposition
is claimed). These results can be generalized to 
constraints on the emission of other light particles that couple to 
spin, as {\em e.g.} in models with additional gauge bosons coupled to the
spins of electrons \cite{Dobrescu}. 

The main set-up of our calculation is given in the next section. Section 3
presents the results for the axi-electric cross sections. Section 4
contains calculations of the expected signal from the solar axion absorption,
and the Appendix provides additional details on atomic calculations.

\section{Theory}

The Hamiltonian for the pseudoscalar axion $a$ interacting with
electrons can be written in two equivalent ways \cite{Pospelov}
(see also Appendix)
\begin{eqnarray}
  \hat H_a = 2\frac{m_e}{f_a} a \bar \psi i \gamma_5 \psi, \\
  \hat H_a = -\frac{\partial_\mu a}{f_a}  \bar \psi \gamma^\mu
  \gamma_5 \psi.
\label{eq:ha}
\end{eqnarray}
where energy scale parameter $f_a$ parameterizes the strength of the
interaction, $m_e$ is electron mass, $a$ is axion field,  $\psi$ is electron
Dirac field. 

Following our previous work we present the cross section of the atomic
ionization by absorbing an axion in a form which contains a dimensionless
function of the axion energy $K(\epsilon_a)$:
\begin{equation}
  \sigma_a(\epsilon_a) = \left(\frac{\epsilon_0}{f_a}\right)^2 \frac{c}{v}
  K(\epsilon_a) a_0^2.
\label{eq:csa}
\end{equation}
where $\epsilon_0$ is an energy scale (in our calculations $\epsilon_0$=1
a.u. = 27.21 eV, but it can also be any other energy unit), $c$ is speed 
of light, $v$ is the axion velocity in the laboratory frame, 
$a_0 = 0.52918 \times 10^{-8}$ cm is Bohr radius, 
$\epsilon_a$ is axion energy.
The function $K(\epsilon_a)$ has no unknown parameters and it is to be found
from numerical calculations. It can be presented in a form
\begin{equation}
  K(\epsilon_a) = \frac{4\pi}{\alpha^2}\frac{1}{\epsilon_a\epsilon_0^2}
  \sum_{L,c,\kappa} (2L+1)\langle \kappa ||\hat H_a||n_c\kappa_c\rangle^2,
\label{eq:k}
\end{equation}
where $c$ is a state in atomic core, $n_c$ and $\kappa_c$ are its principal
and angular quantum numbers, $\kappa$ is an angular quantum number for a state
in the continuum. Summation over $L$ saturates very rapidly, we cut it at
$L_{max}=3$. We use relativistic Hartree-Fock method to calculate electron
wave functions in the core and in the continuum.

The form of the single-electron matrix element depends on the
form of the Hamiltonian for the axion-electron interaction (see Appendix for
details). The first form (see formula (\ref{Eq:TauLRmel}-\ref{Eq:x-section}) in the Appendix) is
simple. However, it often leads to unstable results. This is due to strong
cancellation between the $P_iQ_j$ and $Q_iP_j$ terms in the radial
integral. The cancellation is of the order of $1/(Z\alpha)^2$ which means that
the formula can be reliably used only for heavy atoms (e.g., Xe).

Second form of the single-electron matrix element (see formulas
(\ref{Eq:TauPrimeLRmel}), (\ref{Eq:TauPrimeLRmel1}) and
(\ref{Eq:tauSreduced}) in the Appendix) is more complicated.
However it is more
convenient for the calculations since it gives stable results. In spite of some
numerical problems, comparing calculations with two different expressions is a
valuable test of the calculations. Two forms of the Hamiltonian must give the
same results when exact electron wave functions are used. Since we use
the Hartree-Fock wave functions we can have only approximate agreement between
results. Therefore, comparing the results is not only a test for the computer
code but also a test for the quality of the wave functions used. In our
experience the results agree within 10\% for the cases when first form gives
stable answers. The term ``stable'' means that variation of the axion energy
leads to smooth change in the absorption cross section.  

Note that all formulas in the Appendix are for a closed-shell
atom. However, this is inessential in our case. We consider axion
energies ($\epsilon_a \ge 1$~keV) for which the effect is strongly
dominated by inner closed shells while contribution from open valence
shells is small and can be neglected. This means that the results can
be used for any atom or ion with closed inner shells. They can also be
used for molecules and condensed matter systems since inner atomic
states depend very little on the environment.

In our present calculations the axion absorption cross section depends on its
mass. The only expression which depends on axion mass explicitly is the
expression for the axion wave vector
\begin{equation}
  k_a = \frac{1}{\hbar c}\sqrt{\epsilon_a^2-(m_ac^2)^2}.
\label{eq:ka}
\end{equation}
In an extreme case of heavy axion ($m_ac^2=\epsilon_a$), axion wave vector $k_a=0$
and only term with $L=0$ contributes to the summation in (\ref{eq:k}). This
case was considered in our previous work~\cite{DzuFlaPos}. 

In present work we consider both these cases. First case gives us a test for
the computer code. The results are the same as in our previous
calculations~\cite{DzuFlaPos}. Second case gives new results for axion
absorption cross section by atoms. According to Ref.~\cite{Pospelov} the ratio
of the absorption probabilities for these two extreme cases is equal to 2/3 in the
non-relativistic limit
\begin{equation}
  \frac{\sigma_a(m_a=0)c}{\sigma_a(m_ac^2=\epsilon_a)v}=\frac{2}{3}.
\label{eq:23}
\end{equation}
Below we will discuss relativistic corrections to this formula.

\section{Calculations of the cross sections}

\begin{figure}
\epsfig{figure=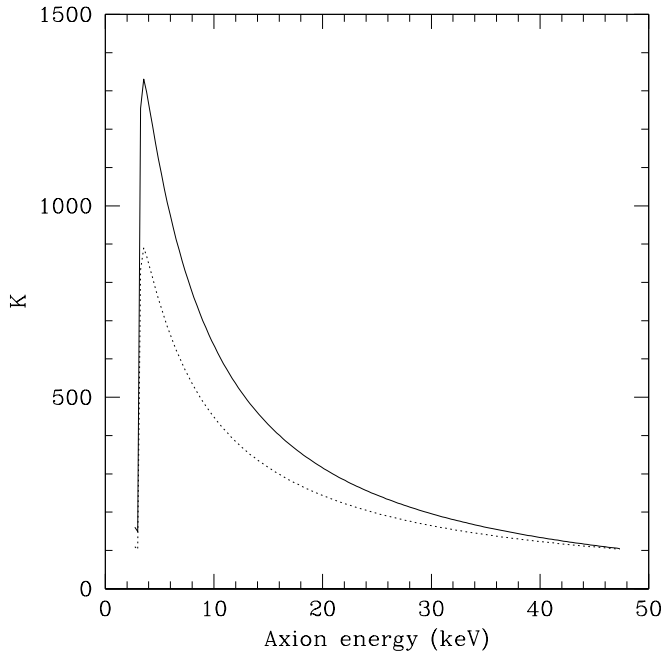,scale=0.45}
\caption{Dimensionless factor $K$ (see formula (\ref{eq:k})) in the
  ionization cross sections of Ar by axion. Solid line - massive
axion ($m_ac^2=\epsilon_a$), dotted line - massless axion ($m_a=0$).}
\label{fig:ar}
\end{figure}

\begin{figure}
\epsfig{figure=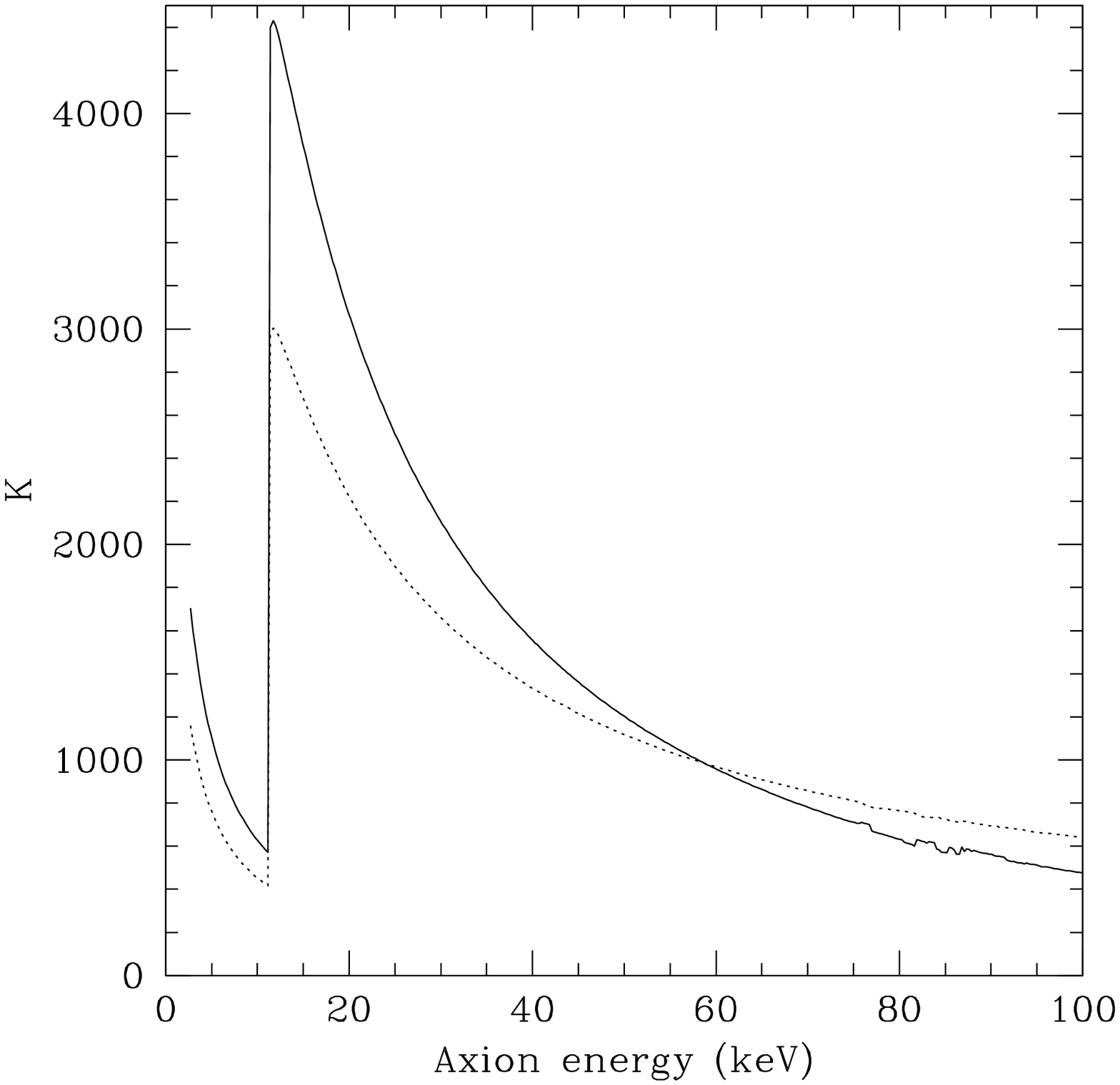,scale=0.45}
\caption{As on Fig.~\ref{fig:ar} but for Ge.}
\label{fig:ge}
\end{figure}

\begin{figure}
\epsfig{figure=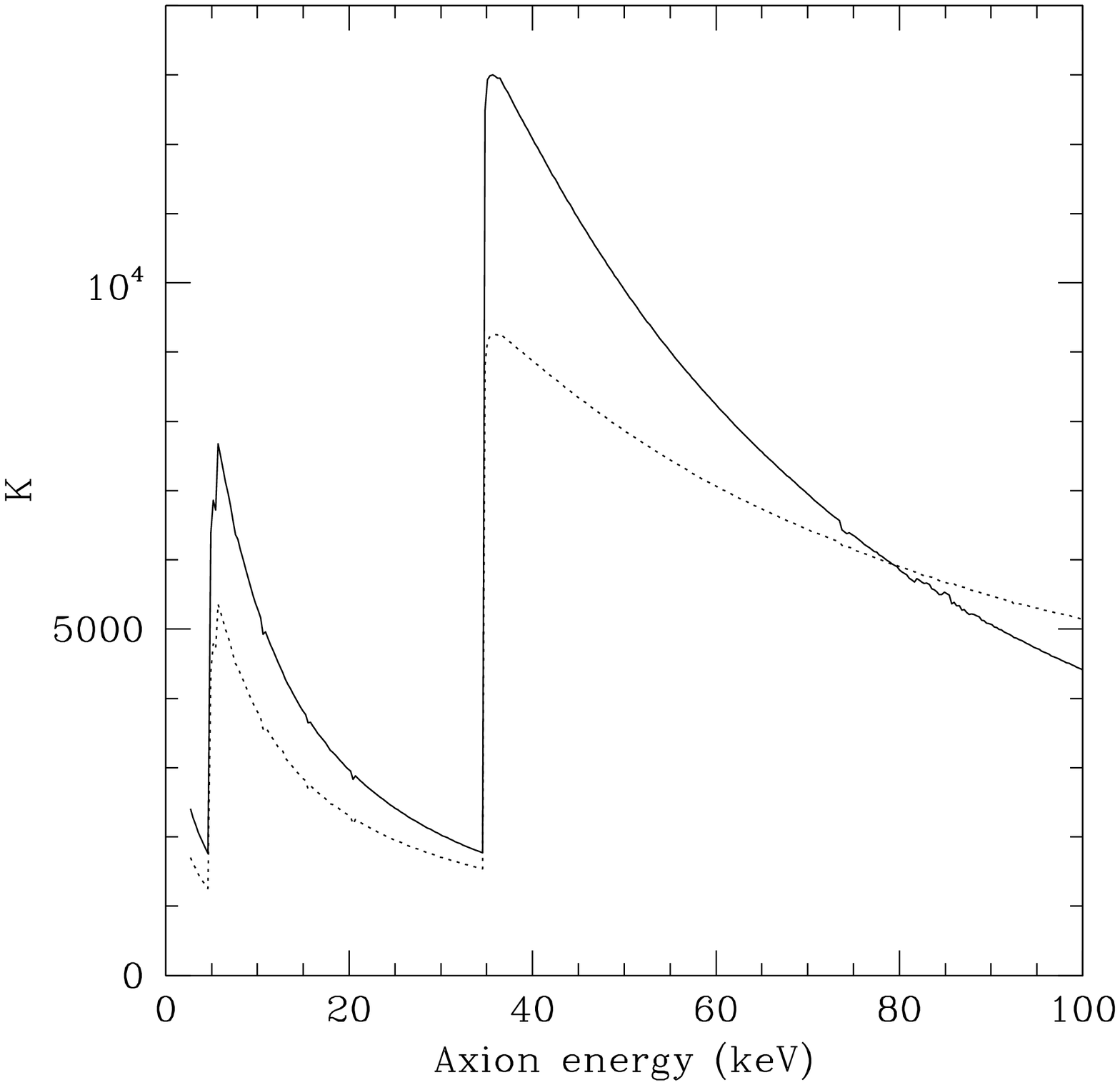,scale=0.45}
\caption{As on Fig.~\ref{fig:ar} but for Xe.}
\label{fig:xe}
\end{figure}

\begin{table} 
\caption{Hartree-Fock energies of the core states of Na, Ar, Ge, I and
  Xe (atomic units, 1 a.u.= 27.21 eV).} 
\label{tab:en}
\begin{tabular}{cccccc}
\hline \hline
Atom  & Na    &  Ar  & Ge  &  I  &  Xe  \\
$Z$   & 11    &  18  & 32  & 53  &  54  \\
\hline
$1s_{1/2}$ & -40.54 &  -119.1 &  -411.0 &  -1225. & -1277. \\ 
$2s_{1/2}$ & -2.805 &  -12.41 &  -53.45 &  -193.0 & -202.4 \\
$2p_{1/2}$ & -1.522 &  -9.631 &  -47.33 &  -180.5 & -189.6 \\
$2p_{3/2}$ & -1.514 &  -9.547 &  -46.14 &  -169.5 & -177.7 \\
$3s_{1/2}$ & -0.1823 & -1.286 &  -7.409 &  -40.52 & -43.01 \\
$3p_{1/2}$ &         & -0.5953 & -5.324 &  -35.34 & -37.66 \\
$3p_{3/2}$ &         & -0.5878 & -5.157 &  -33.21 & -35.32 \\
$3d_{3/2}$ &         &         & -1.616 &  -24.19 & -26.02 \\
$3d_{5/2}$ &         &         & -1.591 &  -23.75 & -25.53 \\
$4s_{1/2}$ &         &         & -0.5687 &  -7.759 & -8.430 \\
$4p_{1/2}$ &         &         & -0.2821 &  -5.868 & -6.452 \\
$4p_{3/2}$ &         &         & -0.2730 &  -5.450 & -5.982 \\
$4d_{3/2}$ &         &         &         &  -2.341 & -2.711 \\
$4d_{5/2}$ &         &         &         &  -2.274 & -2.633 \\
$5s_{1/2}$ &         &         &         & -0.8762 & -1.010 \\
$5p_{1/2}$ &         &         &         & -0.4341 & -0.4925 \\
$5p_{3/2}$ &         &         &         & -0.3903 & -0.4398 \\
\hline \hline
\end{tabular}
\end{table}

Figures \ref{fig:ar},\ref{fig:ge} and \ref{fig:xe} show the results of the
relativistic Hartree-Fock calculations for Ar, Ge and Xe of the dimensionless
function of the axion energy $K(\epsilon_a)$ which 
stands in the expression for the cross 
section of atom ionization by axion (see formula
(\ref{eq:csa})). Many body and relativistic
effects beyond the RHF method are ignored and the final electron state
in the continuum is calculated in the same potential as initial core
state. The accuracy of this approximation is few percents due to
dominating contribution from the inner-most core states $1s$, $2s$ and
$2p$. For these states the many-body effects are small due to strong
nuclear field.  

Solid lines on Figures \ref{fig:ar}, \ref{fig:ge} and \ref{fig:xe} correspond
to the case 
when all axion energy is due to its mass ($m_ac^2 = \epsilon_a$). This is the
same case as was considered in are previous work~\cite{DzuFlaPos}. Dotted line
corresponds to the case of the massless axion ($m_a=0$). One can see that the
ratio of the cross sections is indeed close to 2/3 at low energies (see
formula (\ref{eq:23})). However, the ratio becomes larger at high energies.
For sufficiently high axion energy the absorption cross section for massless
axion becomes larger than those for a massive axion. This is due to the
relativistic effects. We found a formula for the ratio of the cross sections
which fits very well the numerical calculations. The formula reads
\begin{eqnarray}
   R(Z,\epsilon_a) \equiv
   \frac{\sigma_a(m_a=0)}{\sigma_a(m_ac^2=\epsilon_a)}=\frac{2}{3} +
   \nonumber \\ +1.5 \times
  10^{-5} Z^2 + 1.9 \times 10^{-4}\frac{\epsilon_a+\epsilon_{min}}{\epsilon_0}.
\label{eq:23rel}
\end{eqnarray}
Here $Z$ is nuclear charge, $\epsilon_a$ is axion energy, $\epsilon_{min}$ is
the energy of the deepest electron state in the core for which
ionization is possible. Note that all states in the core have negative
energies, therefore ionization threshold corresponds to the condition
$\epsilon_a=-\epsilon_{min}$. Hartree-Fock energies
of all core states for Na, Ar, Ge, I and Xe are presented in
Table~\ref{tab:en}.
The $\epsilon_0$ parameter in (\ref{eq:23rel}) is the energy unit. 
First term on the right-hand side of (\ref{eq:23rel}) corresponds to the
non-relativistic limit; 
second term is the relativistic correction due to core electrons; last term
is the relativistic correction due to the kinetic energy of the escaping
electron. 

In our previous work \cite{DzuFlaPos} we presented an analytical formula which
is an accurate fit of the numerical calculations of the absorption cross
section for the massive axion. The formula can be used for wide range of atoms
and axion energies. 
The formula reads
\begin{eqnarray}
&&K(m_ac^2=\epsilon_a) = K_{1s} + K_{2s} + K_{2p}, \label{eq:kt} \\
K_{1s} &=&
f_1(Z,\epsilon_a+\epsilon_{1s})\frac{384\pi\epsilon_{1s}^4}{(\epsilon_0 Z
  \epsilon_a)^2} \frac{e^{-4\nu_1{\rm arccot}\nu_1}}
{1-e^{-2\pi\nu_1}}, \label{eq:k1s} \\
K_{2s} &=& f_2(Z,\epsilon_a+\epsilon_{2s})\frac{6144\pi
  e_2^3}{\epsilon_0 \epsilon_a^2} 
\left(1+3\frac{e_2}{\epsilon_a}\right) \nonumber \\
&&\times \frac{e^{-4\nu_2{\rm arccot}(\nu_2/2)}}
{1-e^{-2\pi\nu_2}},  \label{eq:k2s} \\
K_{2p} &=& f_2(Z,\epsilon_a+\epsilon_{2p})\frac{12288\pi
  e_3^4}{\epsilon_0 \epsilon_a^3} 
\left(3+8\frac{e_3}{\epsilon_a}\right) \nonumber \\
&&\times \frac{e^{-4\nu_3{\rm arccot}(\nu_3/2)}}{1-e^{-2\pi\nu_3}}, 
 \label{eq:k2p}
\end{eqnarray}
where $\alpha$ is the fine structure constant, $Z$ is nuclear
charge, $\epsilon_a$ is axion energy, $e_2=|\epsilon_{2s}|$,
$e_3=|\epsilon_{2p}|$,  $\nu_1 =
\sqrt{-\epsilon_{1s}/(\epsilon_{1s}+\epsilon_a)}$,  $\nu_2 =
2\sqrt{-\epsilon_{2s}/(\epsilon_{2s}+\epsilon_a)}$,  $\nu_3 =
2\sqrt{-\epsilon_{2p}/(\epsilon_{2p}+\epsilon_a)}$.  
Here $\epsilon_{1s}$, $\epsilon_{2s}$ and  $\epsilon_{2p}$ are the
Hartree-Fock energies of the core states. Hartree-Fock energies of 
the $1s$, $2s$ and $2p_{1/2}$ states of many-electron atoms can be
found using extrapolation formulas:
\begin{eqnarray}
  \frac{\epsilon_{1s}}{\epsilon_0}(Z) &=& -\frac{Z^2-7.49Z+43.39}{2},
\label{eq:1s} \\
  \frac{\epsilon_{2s}}{\epsilon_0}(Z) &=& -0.000753 Z^3-0.028306 Z^2 \nonumber \\
&&-0.066954 Z \ +2.359052, \label{eq:2s} \\
  \frac{\epsilon_{2p}}{\epsilon_0}(Z) &=& -0.000739 Z^3-0.027996 Z^2 \nonumber \\
&&+0.128526 Z \ +1.435129. \label{eq:2p} 
\end{eqnarray}
The functions $f_1(Z)$ and $f_2(Z)$ in
(\ref{eq:k1s},\ref{eq:k2s},\ref{eq:k2p}) are scaling functions:
\begin{eqnarray}
         f_1(Z,\epsilon) &=& (5.368 \times 10^{-7}Z -
         1.17\times 10^{-4})\epsilon/\epsilon_0 \nonumber \\
         &&  - 0.012Z + 1.598 \label{eq:f1} \\
         f_2(Z,\epsilon) &=& (-1.33 \times 10^{-6}Z +
         1.17\times 10^{-4})\epsilon/\epsilon_0 \nonumber \\
         &&  - 0.0156Z + 1.15 \label{eq:f2} 
\end{eqnarray}
To find a cross section for massless axion one should take formula
(\ref{eq:kt}) and multiply it by the factor $R(Z,\epsilon_a)$ given by
(\ref{eq:23rel}). 
Therefore, for the massless axion we also have
the results which cover the same range of atoms and energies as in
Ref.~\cite{DzuFlaPos}.

\section{Solar axion absorption signal}

To calculate the rate of the axio-electric effect caused by solar axions, we 
first address the issue of the total axion flux. Both continuous and line-like emission 
is possible. Here we take into account the emission of solar axions 
due to their couplings to nucleons, to photons and to electrons. 
The easiest case  to address is the nuclear case, as it leads to a 
characteristic $E_a=14.4$ keV emission due the nuclear 
transition of the $ ^{57}$Fe nucleus \cite{Haxton}. The 
solar axion flux was calculated in Ref. \cite{CAST2} (where 
CAST results were also used to constrain it in combination with coupling of
axions to photons). At Earth this flux is given by
\begin{equation}
\Phi_a = 4.5 \times 10^{23} \left(\frac{{\rm 1~ GeV}}{f_{aN}}\right)^2 \times {\rm cm^{-2} s^{-1}},
\end{equation}
where $f_{aN}$ is some effective coupling constant to nucleons that can be related to the 
coupling of axions to quark spins. 
The expected counting rates of argon, germanium and xenon experiments are given by 
\begin{eqnarray}
R_{\rm Ar} \simeq 4 \left( \frac{10^6 {\rm GeV}}{(f_af_{aN})^{1/2}} \right)^4{\rm kg^{-1}day^{-1}},
\\
R_{\rm Ge} \simeq 18 \left( \frac{10^6 {\rm GeV}}{(f_af_{aN})^{1/2}} \right)^4{\rm kg^{-1}day^{-1}},
\\
R_{\rm Xe} \simeq 11 \left( \frac{10^6 {\rm GeV}}{(f_af_{aN})^{1/2}} \right)^4{\rm kg^{-1}day^{-1}},
\end{eqnarray}
where the following values for the $K$-factors are used: 
\begin{eqnarray}
K_{\rm Ar}(14.4~{\rm keV}) &=& 329; \nonumber \\
K_{\rm Ge}(14.4~{\rm keV}) &=& 2746; \nonumber \\
K_{\rm Xe}(14.4~{\rm keV}) &=& 2930. \nonumber
\end{eqnarray}
These rates should provide the sensitivity to $(f_af_{aN})^{1/2}$
in the window between $10^6$ and $10^7$ GeV. Similar strength 
constraints were derived in the recent work \cite{Avignone_recent},
where a $\sim 3\%$ annual modulation of the axion signal was exploited in conjunction 
with DAMA results. (Unlike the signal from WIMP dark matter that is expected to have a
maximum in June, the solar axion signal is minimized in early July.)
We leave it to the experimental collaborations to determine the exact 
upper limits on solar axions ensuing from their results. 

\begin{figure}
\epsfig{figure=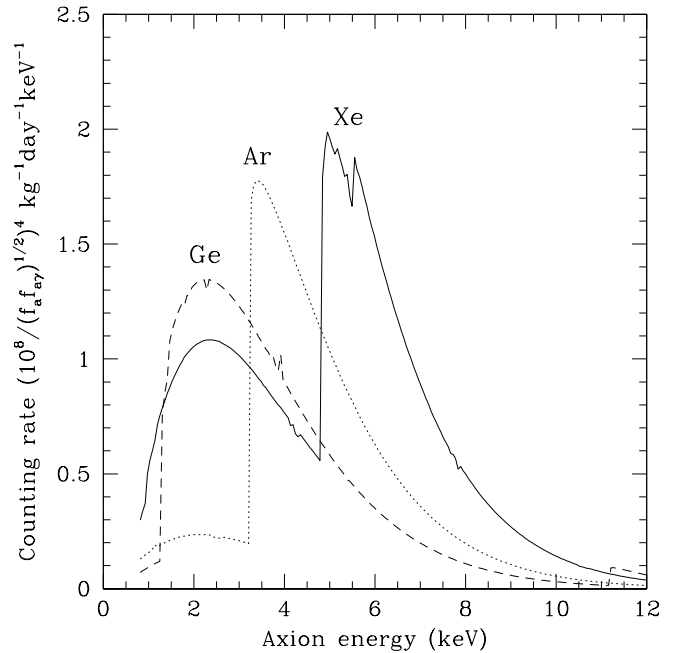,scale=0.45}
\caption{Counting rate for the axio-electric effect for Ar, Ge and Xe
  as a function of axion energy.}
\label{fig:rates}
\end{figure}

If the coupling to photons is not zero, $F_{\mu\nu}\tilde F_{\mu\nu}a/(4f_{a\gamma})$, 
then we can calculate the counting rate, using the axion flux provided in 
Ref.~\cite{CAST}: 
\begin{eqnarray}
&\frac{d\Phi_a}{d\epsilon_a} = 6.02 \times 10^{30} \left(\frac{{\rm 1~
      GeV}}{f_{a\gamma}}\right)^2
\epsilon_a^{2.481}e^{-\frac{\epsilon_a}{1.205}} \label{eq:flux} \\ &\times {\rm cm^{-2}
  \ s^{-1} \ keV^{-1}}. \nonumber
\end{eqnarray}
Counting rate for the axio-electric effect is given by the product of
the calculated absorption cross section and the flux (\ref{eq:flux}).
For $(f_a f_{a\gamma})^{1/2}$ normalized on $10^8$ GeV, we get the 
counting rates plotted in Figure \ref{fig:rates}. 
Integration over axion energy leads to the following total counting rates
\begin{eqnarray}
R_{\rm Ar} \simeq 5.0 \left( \frac{10^8 {\rm GeV}}{(f_af_{a\gamma})^{1/2}} \right)^4{\rm kg^{-1}day^{-1}},
\\
R_{\rm Ge} \simeq 5.2 \left( \frac{10^8 {\rm GeV}}{(f_af_{a\gamma})^{1/2}} \right)^4{\rm kg^{-1}day^{-1}},
\\
R_{\rm Xe} \simeq 8.2 \left( \frac{10^8 {\rm GeV}}{(f_af_{a\gamma})^{1/2}} \right)^4{\rm kg^{-1}day^{-1}}.
\end{eqnarray}

Comparing this to the counting rate of 
the CDMS experiment \cite{CDMS}, one can see that the equivalent of 
$(f_a f_{a\gamma})^{1/2} \sim 10^8$ GeV are being probed, 
as the counting rates in the window from 1.5 to 4 keV 
reach $O(1\rm~kg^{-1}day^{-1}keV^{-1})$. Similar sensitivity is achieved in the 
CoGent experiment \cite{CoGent}.

Finally, the axion flux can be created by the emission of the 
axions due to the same interaction that leads to atomic ionization. 
In this case, however, the production cross section is down 
by additional factor of $E_a^2/m_e^2$ \cite{Pospelov}, and the sensitivity to 
$f_a$ in this case does not exceed $10^6$ GeV. 

\section{Conclusions}

QCD axions represent one of the most well-motivated extensions of the Standard Model. 
Their light mass and small couplings allow them to be produced in the Solar interior and 
escape reaching the Earth. With the proliferation of the low-background searches of 
dark matter, one should also conduct searches of solar axions. In this paper we have calculated 
the cross sections relevant for these searches, improving upon the simple scaling relations
that tie the axio-electric and photo-electric effects. 

Last two years has brought a significant progress in sensitivity to any ionizing effects in 
Germanium in the window from 1 to 10 keV \cite{CDMS,CoGent}. 
Currently, the CoGent experiment has very low backgrounds in the window from 
2 to 4 keV, where the solar axion signal is expected to peak. With acquiring more statistics, the 
sensitivity to the solar axions in this experiment is poised to grow. We also remark at this point that the 
excess of events recorded by CoGent below 1 keV does not fit the expected shape of the 
spectrum from solar axions. Future progress in searching for solar axions may also come from the large scale 
detectors with self-shielding capabilities.

\acknowledgements

This work was supported in part by the Australian Research Council,
and the NSERC of Canada. Research at Perimeter Institute is also supported in part by NSERC 
and by the Province of Ontario through MEDT.

\appendix
\section{Derivation of cross-section for the axio-electric effect}
There are two equivalent expressions
for the Lagrangian describing coupling of pseudoscalar axions of mass $m_a$ to electrons
\begin{align}
H_{a} &  =2\frac{m_{e}}{f_{a}}a~\overline{\psi}i\gamma_{5}\psi,\label{Eq:H}\\
H_{a}^{\prime} &  =-\frac{1}{f_{a}}\left(  \partial_{\mu}a\right)
~\overline{\psi}\gamma^{\mu}\gamma_{5}\psi.\label{Eq:Hp}%
\end{align}
Here $\psi$ is the electronic wave function and $f_{a}$ is a coupling constant.
The axion field $a$ may be represented as
\[
a=N~e^{i\left(  \mathbf{k}\cdot\mathbf{r}-\omega t\right)  }=Ne^{-ik_{\lambda
}x^{\lambda}},
\]
with the dispersion relation%
\[
\hbar\omega=k_{0}=\sqrt{\left(  mc^{2}\right)  ^{2}+\left(  \hbar c\left\vert
\mathbf{k}\right\vert \right)  ^{2}},
\]
and $N$ being the normalization constant. 

We will treat the axio-ionization in
the independent-particle approximation (IPA) for atom. In the IPA, the atomic
many-body wave function is represented by a single Slater determinant built
from single-particle atomic orbitals. Then, as a result of the
axio-ionization, an atomic electron in the initial single-particle orbital
will be ejected into a continuum state. The standard prescription for
evaluating cross-sections due to $H_{a}$ and $H_{a}^{\prime}$ requires
computing  matrix elements of transition operators,%
\begin{align}
T_{a} &  =-e^{i\mathbf{k}\cdot\mathbf{r}}i\gamma_{0}\gamma_{5},\label{Eq:T}\\
T_{a}^{\prime} &  =\frac{1}{2}e^{i\mathbf{k}\cdot\mathbf{r}}\left(  ik_{\mu
}\right)  \gamma_{0}\gamma^{\mu}\gamma_{5}.\label{Eq:Tp}%
\end{align}
In the last formula $\gamma_{0}k_{\mu}\gamma^{\mu}\gamma_{5}=\gamma_{5}%
k_{0}-\mathbf{k}\cdot\mathbf{\sigma}$.
Formal equivalence of matrix elements from pseudoscalar and axial-vector forms of the 
interactions was demonstrated in Ref. \cite{Pospelov} with the use of the single-electron Dirac equation 
with arbitrary potential. 

Since atoms are spherically-symmetric, we employ the standard machinery of
the angular momentum algebra~\cite{VarMosKhe88} and use the partial wave expansion for evaluating
cross-sections. In particular, at large values of electronic coordinate the
continuum orbital has to go over to a sum of an incoming spherical and plane
waves~\cite{Tay83}. Scattering wave function satisfying this boundary condition
may be decomposed in partial waves
\begin{equation}
w_{\mathbf{p}\lambda}=N_{p}\sum_{\kappa m}\left(  \Omega_{\kappa m}^{\dagger
}\left(  \widehat{p}\right)  \,\chi_{\lambda}\right)  \,\,i^{l-1}%
e^{-i\delta_{\kappa}}\,w_{\kappa m}\left(  \mathbf{r}\right)
\,.\label{Eqn_cont}%
\end{equation}
Here $\Omega_{\kappa m}$ is a spherical spinor, $\chi_{\sigma}$ is
a two-component spinor describing spin-polarization of the photoelectron, the
relativistic angular quantum number $\kappa=(l-j)(2j+1)$ is expressed in terms
of the total $j$ and orbital $\ l$ angular momenta, and $\delta_{\kappa}$ is a
scattering phase shift. For box-normalized solutions ($V$ is the volume of the box, $p$ and $E$ are
the momentum and the energy of the electron, and $\ \alpha_{fs}$ is the
fine-structure constant)%
\[
N_{p}=\left(  \frac{\left(  2\pi\right)  ^{3}}{\alpha_{fs}\,E\,p\,V}\right)
^{1/2} \, .
\]

Wave function $w_{\kappa m}\left(  \mathbf{r}\right)
$ may be expressed in terms of the large ($S_{\kappa}$) and small ($T_{\kappa
}$) components satisfying the radial Dirac equations
\begin{equation}
w_{\kappa m}\left(  \mathbf{r}\right)  =\frac{1}{r}\left(
\begin{array}
[c]{c}%
iS_{\kappa}(r)\,\Omega_{\kappa m}(\hat{r})\\
T_{\kappa}(r)\,\Omega_{-\kappa m}(\hat{r})
\end{array}
\right)  \,.
\end{equation}
For bound-state orbitals, the parameterization reads%
\begin{equation}
|n_{b}\kappa_{b}m_{b}\rangle=\frac{1}{r}\left(
\begin{array}
[c]{c}%
iP_{n\kappa}(r)\ \Omega_{\kappa m}(\hat{r})\\
Q_{n\kappa}(r)\ \Omega_{-\kappa m}(\hat{r})
\end{array}
\right)  \,.\label{Eq:BiSpinorSpher}%
\end{equation}
Axio-ionization cross-sections are proportional to the square of transition
amplitudes. Averaging it over all possible spin polarizations $\lambda$,
magnetic quantum numbers $m_{b}$ and $m$ and integrating over the
directions of the ejected electron momentum, we find
$
\sum_{\lambda\kappa m\,m_{b}\kappa_{b}}\int|\langle w_{\mathbf{p}\lambda
}|T_{a}|n_{b}\kappa_{b}m_{b}\rangle|^{2}d\Omega_{p}=N_{p}^{2}\sum_{\kappa
m\,m_{b}}\,\left\vert \langle w_{\kappa m}|T_{a}|n_{b}\kappa_{b}m_{b}%
\rangle\right\vert ^{2}.
$
The same result holds for the averaged square of $T_{a}^{\prime}$ operator.

The involved matrix element $\langle w_{\kappa m}|T_{a}|n_{b}\kappa_{b}%
m_{b}\rangle$ is between the electronic states of definite angular momenta and
parity. For simplifying summations over magnetic quantum numbers, we
expand the transition operators into irreducible tensor operators
(ITO) and then apply the Wigner-Eckart theorem. 

We start with the simpler case of operator (\ref{Eq:T}). 
We employ the conventional expansion
\[
e^{i\mathbf{k}\cdot\mathbf{r}}=\sum_{LM}\left[  L\right]  i^{L}j_{L}\left(
kr\right)  C_{LM}^{\ast}\left(  \hat{k}\right)  C_{LM}\left(  \hat{r}\right)
\, ,
\]
where $C_{LM}$ are the normalized spherical harmonics~\cite{VarMosKhe88} and $\left[
L\right]  =2L+1$.  Then we reexpress the transition operator as%

\begin{equation}
T_{a}=-e^{i\mathbf{k}\cdot\mathbf{r}}i\gamma_{0}\gamma_{5}=\sum_{LM}%
i^{L}~\left[  L\right]  C_{LM}^{\ast}\left(  \hat{k}\right)  \tau
_{LM}(\mathbf{r}).\label{Eq:TaMultipoleExpansion}%
\end{equation}
Here the operators
\[
\tau_{LM}(\mathbf{r})=-~i\gamma_{0}\gamma_{5}~j_{L}\left(  kr\right)
C_{LM}\left(  \hat{r}\right)
\]
are ITOs of rank $L.$ A matrix element evaluated between two atomic orbitals
reads
\begin{widetext}
\[
\left(  \tau_{LM}\right)  _{ij}=-\langle\kappa_{i}m_{i}|C_{LM}|-\kappa
_{j}m_{j}\rangle\left(  \int_{0}^{\infty}j_{L}\left(  kr\right)  \left[
P_{n_{i}\kappa_{i}}(r)Q_{n_{j}\kappa_{j}}(r)+Q_{n_{i}\kappa_{i}}%
(r)\ P_{n_{j}\kappa_{j}}(r)\right]  dr\right)
\]
The selection rules for matrix elements of the C-tensor require that
$\left\vert j_{i}-j_{j}\right\vert \leq L\leq j_{i}+j_{j}$ and $L+l_{i}%
+l_{j}=\mathrm{odd}$. For example, for $L=0$ the multipolar operator is pseudoscalar:
the $\tau_{00}$ operator drives $s_{1/2}\rightarrow p_{1/2}$ transitions.
Reduced matrix element
\begin{equation}
\langle n_{i}\kappa_{i}||\tau_{L}||n_{j}\kappa_{j}\rangle=-\langle\kappa
_{i}||C_{L}||-\kappa_{j}\rangle\left(  \int_{0}^{\infty}j_{L}\left(
kr\right)  \left[  P_{n_{i}\kappa_{i}}(r)Q_{n_{j}\kappa_{j}}(r)+Q_{n_{i}%
\kappa_{i}}(r)\ P_{n_{j}\kappa_{j}}(r)\right]  dr\right)  .\label{Eq:TauLRmel}%
\end{equation}

To evaluate the cross-section we fix the coordinate system in such a way that
the axion propagates along the $z$-axis. Then in
Eq.(\ref{Eq:TaMultipoleExpansion}),
\begin{equation}
C_{LM}^{\ast}\left(  \hat{k}\right)  =\,\delta_{M0}\text{ and }T_{a}=\sum
_{L}i^{L}[L]~\tau_{LM=0}(\mathbf{r}).
\end{equation}
Further%
\[
\sum_{\lambda\kappa m\,}\sum_{n_{b}\kappa_{b}m_{b}}\int|\langle w_{\mathbf{p}%
\lambda}|T_{a}|n_{b}\kappa_{b}m_{b}\rangle|^{2}d\Omega_{p}=N_{p}^{2}%
\sum_{n_{b}\kappa_{b}\kappa L}\left(  2L+1\right)  \left(  \langle\kappa
||\tau_{L}||n_{b}\kappa_{b}\rangle\right)  ^{2}.
\]
Finally,%
\begin{equation}
\sigma=\frac{c}{v}\frac{4\pi}{\alpha_{fs}^{2}}\frac{1}{f_{a}^{2}}\frac
{1}{\varepsilon_{a}}\times\sum_{n_{b}\kappa_{b}\kappa L}~\left(  2L+1\right)
~\left(  \langle\kappa||\tau_{L}||n_{b}\kappa_{b}\rangle\right)
^{2}.\label{Eq:x-section}%
\end{equation}

Derivation of the axio-ionization cross-section for the alternative form of
the coupling $H_{a}^{\prime}$ \ is more complicated. We start from the
multipole expansion of the $T_{a}^{\prime}$ operator,%
\begin{equation}
T_{a}^{\prime}=\frac{1}{2}i\left(  \gamma_{5}k_{0}-\mathbf{k}\cdot
\mathbf{\sigma}\right)  e^{i\mathbf{k}\cdot\mathbf{r}}=\sum_{LM}i^{L}~\left[
L\right]  C_{LM}^{\ast}\left(  \hat{k}\right)  \tau_{LM}^{\prime}%
(\mathbf{r}).\label{Eq:MultExpTp}%
\end{equation}
Because the angular dependence of this expansion is the same as in
Eq.(\ref{Eq:TaMultipoleExpansion}), the expression for \ the cross-section
remains the same as in the $H_{a}$ case, Eq.(\ref{Eq:x-section}), with the
substitution $\tau_{L}\rightarrow\tau_{L}^{\prime}.$

The multipolar tensors $\tau_{LM}^{\prime}$ may be derived by inverting the
expansion (\ref{Eq:MultExpTp})%
\begin{align*}
\tau_{LM}^{\prime}(\mathbf{r}) &  =\frac{i^{-L}}{4\pi}\int d\Omega_{k}%
C_{LM}\left(  \hat{k}\right)  \frac{1}{2}i\left(  \gamma_{5}k_{0}%
-\mathbf{k}\cdot\mathbf{\sigma}\right)  e^{i\mathbf{k}\cdot\mathbf{r}}=\\
&  =\sum_{L^{\prime}M^{\prime}}i^{L^{\prime}-L}\left[  L^{\prime}\right]
j_{L^{\prime}}\left(  kr\right)  C_{L^{\prime}M^{\prime}}\left(  \hat
{r}\right)  \frac{1}{4\pi}\int d\Omega_{k}C_{LM}\left(  \hat{k}\right)
\frac{1}{2}i\left(  \gamma_{5}k_{0}-\mathbf{k}\cdot\mathbf{\sigma}\right)
C_{L^{\prime}M^{\prime}}^{\ast}\left(  \hat{k}\right)  .
\end{align*}
The two contributions to the integral are
\begin{equation}
\frac{1}{4\pi}\int d\Omega_{k}C_{LM}\left(  \hat{k}\right)  C_{L^{\prime
}M^{\prime}}^{\ast}\left(  \hat{k}\right)  \frac{1}{2}i\left(  \gamma_{5}%
k_{0}\right)  =\frac{1}{\left[  L\right]  }\frac{1}{2}i\left(  \gamma_{5}%
k_{0}\right)  \delta_{LL^{\prime}}\delta_{MM^{\prime}}%
\end{equation}
and%
\begin{equation}
\frac{1}{4\pi}\int d\Omega_{k}C_{LM}\left(  \hat{k}\right)  C_{L^{\prime
}M^{\prime}}^{\ast}\left(  \hat{k}\right)  \frac{1}{2}i\left(  -\mathbf{k}%
\cdot\mathbf{\sigma}\right)  =-\frac{1}{2}\left\vert \mathbf{k}\right\vert
i\left(  \mathbf{\sigma}\cdot\mathbf{a}_{L^{\prime}M^{\prime};LM}\right)  ,
\end{equation}
where the components of a vector object $\mathbf{a}_{L^{\prime}M^{\prime}%
;LM}$
\begin{equation}
\left(  \mathbf{a}_{L^{\prime}M^{\prime};LM}\right)  _{\lambda}=\frac{1}{4\pi
}\int d\Omega_{k}C_{L^{\prime}M^{\prime}}^{\ast}\left(  \hat{k}\right)
\hat{k}_{\lambda}C_{LM}\left(  \hat{k}\right)  =\frac{1}{\sqrt{\left[
L\right]  \left[  L^{\prime}\right]  }}\langle L^{\prime}M^{\prime
}|C_{1\lambda}|LM\rangle.
\end{equation}
The resulting expression reads%
\[
\tau_{LM}^{\prime}(\mathbf{r})=\tau t_{LM}^{\prime}(\mathbf{r})+\tau
s_{LM}^{\prime}(\mathbf{r})=j_{L}\left(  kr\right)  C_{LM^{\prime}}\left(
\hat{r}\right)  \frac{1}{2}i\left(  \gamma_{5}k_{0}\right)  -\frac{1}%
{2}\left\vert \mathbf{k}\right\vert \sum_{L^{\prime}M^{\prime}}i^{L^{\prime
}-L+1}\left[  L^{\prime}\right]  j_{L^{\prime}}\left(  kr\right)
C_{L^{\prime}M^{\prime}}\left(  \hat{r}\right)  \left(  \mathbf{\sigma}%
\cdot\mathbf{a}_{L^{\prime}M^{\prime};LM}\right)  ,
\]
where we split the operator into the time- and space-like contributions. Below
we tabulate reduced matrix elements of the $\tau_{LM}^{\prime}$ ITO %
\begin{equation}
\langle n_{i}\kappa_{i}||\tau_{L}^{\prime}||n_{j}\kappa_{j}\rangle=\langle
n_{i}\kappa_{i}||\tau t_{L}^{\prime}||n_{j}\kappa_{j}\rangle+\langle
n_{i}\kappa_{i}||\tau s_{L}^{\prime}||n_{j}\kappa_{j}\rangle
,\label{Eq:TauPrimeLRmel}%
\end{equation}%
\begin{equation}
\langle n_{i}\kappa_{i}||\tau t_{L}^{\prime}||n_{j}\kappa_{j}\rangle=\frac
{1}{2}k_{0}\langle\kappa_{i}||C_{L}||-\kappa_{j}\rangle\left(  \int%
_{0}^{\infty}j_{L}\left(  kr\right)  \left[  P_{n_{i}\kappa_{i}}%
(r)Q_{n_{j}\kappa_{j}}(r)-Q_{n_{i}\kappa_{i}}(r)\ P_{n_{j}\kappa_{j}%
}(r)\right]  dr\right)  ,\label{Eq:TauPrimeLRmel1}%
\end{equation}%
\begin{align}
\langle n_{i}\kappa_{i}||\tau s_{L}^{\prime}||n_{j}\kappa_{j}\rangle &
=-\frac{1}{2}\left\vert \mathbf{k}\right\vert \sum_{L^{\prime}%
=L-1,L+1;L^{\prime}\geq0}i^{L^{\prime}-L+1}\langle\kappa_{i}||A_{L^{\prime}%
}||\kappa_{j}\rangle\left[  L^{\prime}\right]  \times\label{Eq:tauSreduced}\\
&  \left(  \int_{0}^{\infty}j_{L^{\prime}}\left(  kr\right)  \left[
P_{n_{i}\kappa_{i}}(r)P_{n_{j}\kappa_{j}}(r)+Q_{n_{i}\kappa_{i}}%
(r)Q_{n_{j}\kappa_{j}}(r)\ \right]  dr\right)  , \nonumber
\end{align}
with%
\begin{align*}
\langle\kappa^{\prime}||A_{L^{\prime}}||\kappa\rangle &  =\langle
\kappa^{\prime}||\sum_{M^{\prime}}C_{L^{\prime}M^{\prime}}\left(  \hat
{r}\right)  \left(  \mathbf{\sigma}\cdot\mathbf{a}_{L^{\prime}M^{\prime}%
;LM}\right)  ||\kappa\rangle=\\
&  \frac{\left[  j,j^{\prime},l,l^{\prime}\right]  ^{1/2}}{\left[
L,L^{\prime}\right]  ^{1/2}}\sqrt{6}\left(  -1\right)  ^{j^{\prime}%
+j}\;(-1)^{L^{\prime}+l}\;\left(
\begin{array}
[c]{ccc}%
l^{\prime} & L^{\prime} & l\\
0 & 0 & 0
\end{array}
\right)  \;\left(
\begin{array}
[c]{ccc}%
L^{\prime} & 1 & L\\
0 & 0 & 0
\end{array}
\right)  \left\{
\begin{tabular}
[c]{lll}%
$1$ & $L^{\prime}$ & $L$\\
$1/2$ & $l^{\prime}$ & $j^{\prime}$\\
$1/2$ & $l$ & $j$%
\end{tabular}
\ \ \right\}  .
\end{align*}
Here the notation $\left[  J_{1},...,J_{n}\right]  =\left(  2J_{1}+1\right)
...\left(  2J_{n}+1\right)  $. The two-row quantities are the 3j-symbols and
the 3x3 matrix in the curly brackets is the 9j-symbol. Notice that the phase
$i^{L^{\prime}-L+1}$ entering Eq.(\ref{Eq:tauSreduced}) is either +1 or -1,
i.e., the entire expression is real. Selection rules for both time- and
space-like contributions are the same as in the case of the $\tau_{LM}$
multipoles: $\left\vert j_{i}-j_{j}\right\vert \leq L\leq j_{i}+j_{j}$ and
$L+l_{i}+l_{j}=\mathrm{odd}$.

To summarize, the cross-section for axio-ionization is given by Eq.~(\ref{Eq:x-section}),
with reduced matrix elements given by Eqs.~(\ref{Eq:TauLRmel}) and (\ref{Eq:TauPrimeLRmel}).
Notice that the derived expressions remain valid for arbitrary large values of parameter $kr$, i.e., even when the usual dipole approximation ($e^{ikr} \approx 1$) breaks down.
\end{widetext}

\end{document}